\documentstyle[11pt]{article}

\newcommand{\be}{\begin{equation}}
\newcommand{\ee}{\end{equation}}
\newcommand{\ba}{\begin{array}}
\newcommand{\ea}{\end{array}}
\newcommand{\bc}{\begin{center}}
\newcommand{\ec}{\end{center}}
\newcommand{\bi}{\begin{itemize}}
\newcommand{\ei}{\end{itemize}}

\newcommand{\disregard}[1]{{}}

\def\bild#1\over#2{\mathrel{\mathop{\kern0pt #1}\limits_{#2}}}

\begin{document}
\centerline {{\bf   CALOGERO MODELS FOR DISTINGUISHABLE PARTICLES \rm}}
\vskip 2cm
\centerline{{\bf Cyril FURTLEHNER\rm } and
{\bf St\a'ephane OUVRY \rm }\footnote{\it  and
LPTPE, Tour 16, Universit\'e Paris  6 / electronic e-mail:
OUVRY@FRCPN11}}
\vskip 1cm
\centerline{{Division de Physique Th\'eorique \footnote{\it Unit\a'e de
Recherche  des
Universit\a'es Paris 11 et Paris 6 associ\a'ee au CNRS},  IPN,
  Orsay Fr-91406}}
\vskip 3cm
{\bf Abstract :} Motivated
by topological bidimensional
quantum models for distinguishable particles, and
by Haldane's definition of mutual statistics for
different species of particles,
we propose a new class of one-dimensional $1/r_{ij}^2$ Calogero model with
coupling constants $g_{ij}$ depending on the labels of the particles. We
solve the groundstate problem, and show how to build some classes of excited
states.

\vskip 3cm

IPNO/TH 94-02

May 1994

\vfill\eject

1. INTRODUCTION

It has been known for some time [1] that there is a
correspondance between the $1/r^2$ $N$-body Calogero model [2] in an
harmonic well and the $N$-anyon
model [3] projected in the
lowest Landau level of an external  magnetic field.
This correspondance can be seen at the level of the spectrum -an
explicit mapping between both models is however still missing-, and also
at the level of thermodynamical quantities, such as the equation of
state of both systems [4], [5].  Actually, one finds thermodynamical
quantities
[6]
appropriate for describing a system of particles with fractional
statistics \`a la Haldane [7]. This last result suggests in particular that
Haldane's statistics is essentially a one dimensional projection of
the anyonic statistics, which is a bidimensional concept.

A point of particular interest in Haldane's
framework
is the possibility to have particles of different species,
with mutual statistical  coupling parameter depending on the species
coupled. This is to be related
to the generalisation of the
anyon model to a model of bidimensional distinguishable particles
[8],
with mutual topological interactions (and possibly an external magnetic
field). Here, one
starts
from a generalized Chern-Simons Lagrangian with $N$ independant gauge
fields coupled to $N$ particles via an $N$x$N$ coupling matrix. This model
is truly a "toplogical quantum model", which narrows down to the usual anyon
model if one insists on particles indistinguishability and on related
statistical considerations.

The existence of such models for different species of particles strongly
suggests, through a line of reasonning identical to the one used above, the
possibility of generalising the
Calogero model to a model of distinguishable particles with $1/r_{ij}^2$
couplings depending on the labelling of the particles $\{i,j\}$ coupled.
It might
then happen that some sort of
mapping exists between this new type of one-dimensional models and
the topological bidimensional quantum model in a strong magnetic field.

In this short note we propose a possible definition of  such a
generalisation of the Calogero model.  Far from being
able to find the desired mapping, if any,
we simply exhibit the groundstate and show how
to find some classes of excited
states. The problem of
the integrability of this new model remains entire.

2. THE CALOGERO MODEL : A REMINDER

  The usual Calogero model [2] is defined by the Hamiltonian

\be H={1\over 2}[\sum_{i=1}^N -{d_i}^2+{x_i}^2 +
\sum_{j\ne i}{g \over {{x_{ij}}^2}}]\ee
where one has considered particles of unit mass in an harmonic well of unit
strength. The eigenstates must have a
well-defined asymptotic behavior to have an Hilbert
space stable under the action of the
Hamiltonian
\be\psi \sim _{x_{ij}=0}x_{ij}^\nu\ee
with $\nu$ solution of
$g=\nu (\nu -1)$.

Since one wants a well defined groundstate, $g$ has to be greater than $-1/4$.
Following standard arguments [2],
one specifies
\be\nu ={1\over2}+{{\sqrt{1+4g}}\over2}\ee
{}From this particular  asymptotic behavior ($\nu$ positive), one concludes
that the particles are
umpenetrable (the current also vanishes at coinciding points).

To proceed further one can perform [1]
 on the Hilbert space the non unitarity transformation
\be \psi =\prod _{j<i}|x_i-x_j|^\nu \tilde\psi \ee
This transformation explicitely encodes the asymptotic behaviour given
in eq.(2). One gets a transformed Calogero Hamiltonian
\be \tilde H ={1\over2}
[\sum_{i=1}^N-d_i^2+x_i^2-\sum_{j\ne i}{\nu \over x_{ij}}(d_i-d_j)]\ee
Calogero [2] has shown that
eigenstates of this Hamiltonian have to be
completely symetric.

The very fact that a particular asymptotic behaviour
is imposed on the eigenstates at coinciding points implies
that different particles configurations are
disconnected. This means  that the Hamiltonian is
still hermitian if restricted on  a Hilbert
space corresponding to a given particle configuration, i.e.
to a given ordering of  particles  (in
other words the particles are unpenetrable) . Thus,
an eigenstate on a given configuration can be
used to define an eigenstate on a different configuration
by simply
multiplying the  eigenstate by a constant which depends on the
desired statistics
\be \psi (Px)=\epsilon _P \psi (x)\ee
where P is the permutation between both configurations.
For example, $\epsilon_P=1$
for a bosonic statistics, and  for
a fermionic statistics, $\epsilon_P$ is the sign of the permutation.
It is however clear   that the information is entirely contained in a given
configuration.

Now,  if one decides to restrict oneself to  the
Hilbert space of totally symetric eigenstates, the
transformed Calogero Hamiltonian coincides with [1]
\be \tilde H={1\over2}\sum_{i=1}^N(a_i^-a_i^++a_i^+a_i^-)\ee
where the creation and annihilation operators have been defined  as
\be a_i^\pm ={1\over\sqrt2}(x_i\mp D_i)\ee
with
\be D_i=d_i+\sum_{j\ne i}{\nu \over x_{ij}}(1-K_{ij})\ee
The commutation rules are
\be [D_i,x_j]=\delta_{ij}(1+\sum_{k=1}^N\nu K_{ik})-\nu K_{ij}\ee
\be [D_i,D_j]=0\ee
\be[a_i^\pm ,a_j^\pm]=0\ee
\be[a_i^-, a_j^+]=\delta_{ij}(1+\sum_{k=1}^N\nu K_{ik})-\nu K_{ij}\ee
\be[\tilde H,{a_i^\pm}^n]=\pm n a_i^\pm\ee
This algebra entirely describes the Calogero model.

As a remark, if one had rather decided to use antisymetric eigenstates,
one would have
defined
\be D_i=d_i+\sum_{j\ne }{\nu \over x_{ij}}(1+K_{ij})\ee
and have applied  the same procedure but now
on a totally antisymetric space.

Or, following [1],  one could as
well have redefined
differently the eigenstates
\be\psi =\prod _{j<i}|x_i-x_j|^{-\nu} \tilde \psi \ee
and then constructed  the same algebra as above but with
\be D_i=d_i-\sum_{j\ne i}{\nu \over x_{ij}}(1-K_{ij})\ee
(and built the eigenstate with the new creation operators on the antisymetric
subspace). In any case, one stresses again that
 all the information is contained in the totally
symetric space.

3. GENERALISATION TO A SYSTEM OF NON IDENTICAL PARTICLES

What has just been said above can be used
to generalise the Calogero model to a system of non identical
particles. The main ingredients or building blocks of the construction
should still be the
asymptotic behavior at coinciding points, and the
algebra  of creation-annihilation operators once this asymptotic
behavior has been extracted. However, in a situation where the particles
are no more identical, once a configuration is given,
there is obviously no way to restrict oneself to
totally symetric eigenstates.

We define the   Hamiltonian
\be H={1\over2}[\sum_{i=1}^N-d_i^2+x_i^2+\sum_{j\ne i}{g_{ij}\over{x_{ij}^2}}
+\sum_{i,j,k\ne }{\nu _{ij}\nu _{ik}\over{x_{ij}x_{ik}}}]\ee
with $g_{ij}=\nu _{ij}(\nu_{ij}-1)$ (again the $g_{ij}$'s have to be greater
than $-1/4$).
One remarks that

i) when all the $\nu_{ij}$'s are equal, this model narrows down
to the usual Calogero model (1)

ii) the occurence of 3-body interactions is
quite reminiscent of what happens in the anyon model [3]

iii) the eigenstates have the same type of asymptotic behavior
as in the usual Calogero model
$\psi \sim _{x_{ij}=0}x_{ij}^{\nu_{ij}} $ with
$\nu_{ij} ={1\over2}+{\sqrt{1+4g_{ij}}\over2}$.

The non unitary transformation
\be\psi =\prod _{j<i}x_{ij}^{\nu _{ij}}\tilde \psi \ee
leads to the Hamiltonian
\be\tilde H ={1\over2}[\sum_{i=1}^N-d_i^2+x_i^2
-\sum_{j\ne i}{\nu_{ij} \over x_{ij}}(d_i-d_j)]\ee
where the 2 an 3-body interactions have simply disappeared. Again, this is
quite
reminiscent of the anyon model where the 3-body interactions can be erased,
altogether with the singular  2-body interactions (singular in
perturbation theory), by
the non unitary transformation $\prod_{i<j}r_{ij}^{|\alpha\vert}$, where
$\alpha$ is the statistical parameter [9].

The  just obtained non hermitian
Hamiltonian is quite analogous to the one encountered in the standard
Calogero case.
Following the same procedure as above, we define
the covariant derivative
\be D_i=d_i+\sum_{j\ne i}{\nu_{ij}  \over x_{ij}}(1-K_{ij})\ee
Its commutator with $x_i$ becomes
\be [D_i,x_j]=\delta_{ij}(1+\sum_{k=1}^N\nu _{ik}K_{ik})-\nu
_{ij}K_{ij}\ee
The main difference is that $[D_i,D_j]$ is now
nontrivial but  happens to vanish
on   totally symetric eigenstates. Thus, one can easily
construct some classes of eigenstates in this
particular space. Indeed, for totally symetric functions,
the Hamiltonian can
be written as
\be \tilde H={1\over2}\sum_{i=1}^N(a_i^-a_i^++a_i^+a_i^-)\ee
with the same
definition for
the $a_i$'s as above. On this symmetric space,
 the commutation relations reduce to
\be [a_i^\pm ,a_j^\pm]=0\ee
\be [a_i^- a_j^+]=\delta_{ij}(1+\sum_{k=1}^N\nu _{ik})-\nu _{ij}\ee
The solution of
\be a_i^-\tilde\psi  =0\ee
is $\tilde\psi _0=\exp  (-\sum_{i=0}^N{{x_i^2}\over2})$.
It is the groundstate since it has has no node. Its
energy is simply $E_0={N\over2}+
{1\over2}\sum_{i\ne j}\nu  _{ij}$.

In analogy with the standard Calogero system, we might also consider the
eigenstates
\be \tilde \psi  _n=(\sum_{i=0}^Na_i^+)^n\tilde\psi  _0\ee
The relation $a_i^-\tilde\psi _n=n\tilde\psi   _{n-1}$ makes it easy to see
that
$\tilde{H}\tilde\psi _n=(E_0+n)\tilde\psi_{n}$, i.e. $\tilde \psi_n$ is
an eigenstate of $\tilde H$ with energy $E_0+n$.
Another directly  generalised eigenstate of the Calogero model is
\be \tilde\psi  _{21}=(\sum_{i=0}^N{a_i^+}^2)\tilde\psi _0
=\sum  _{i=1}^N[2{x_i}^2-(1+\sum_{l\ne  i}\nu   _{il})]\tilde\psi  _0\ee
with $E_0+2$ for eigenvalue.
Actually it is easy to see that any eigenstate of the form
\be \tilde\psi (x) =P_n(x)\tilde\psi_0(x)\ee
with $P_n$ a polynomial of degree $n$ in the variables $x_i$ has
necesseraly $E_0+n$ for eigenvalue since
\be \tilde H \tilde\psi
(x)=[(E_0+n)P_n(x)+Q_{n-2}(x)]\tilde\psi_0(x).\ee
Clearly, $Q_{n-2}$, a polynomial of degree $n-2$,
has to vanish for $\tilde\psi$ to be an
eigenstate of energy $E_0+n$.

We could go on and consider
for example the state
\be \tilde\psi  _{31}=(\sum_{i=0}^N{a_i^+}^3)\tilde\psi   _0
=\sqrt2\sum_{i=1}^N[2{x_i}^3-3x_i(1+\sum_{l\ne  i}\nu _{il})]\tilde\psi_0\ee
However, it is not an eigenstate in the present case
since
\be \tilde H\tilde\psi_{31}=(E_0+3)\tilde\psi_{31} +
{3\over\sqrt2}\sum_{j\ne i}{\nu_{ij}\over x_{ij}}
                     (\sum_{l\ne i}\nu_{il}-\sum_{l\ne j}
\nu_{jl})\tilde\psi_0 \ee
Clearly, the last term in the right hand side of (32)
vanishes if and only if all the $\nu_{ij}$'s are equal.

To close this analysis, let us consider the subspace
of the totally symetric space stable under
the action of $\tilde H  $. It is defined by
\be 0=\sum_{j\ne k,k+1}(\nu  _{jk}-\nu  _{jk+1})({1\over{x_{jk}}}(d_j-d_k)-
{1\over{x_{jk+1}}}(d_j-d_{k+1}))\ee
for $k=1,\ldots,N-1$.
On this particular subspace $\tilde H  $ coincides with the usual
Calogero Hamiltonian
with $\nu  ={2\over{N(N-1)}}\sum_{i\ne j}\nu  _{ij}$. Thus,
projecting the Calogero eigenstates on this subspace, gives all the
eigenstates of the model in this subspace.
But it certainly does not provide the entire set of eigenstates, simply because
most of the eigenstates do not have a
symetrical analytic form, as already stressed above. This is obviously due
to $\tilde H$ not being anymore  symetrical, contrary to
what happens in the standard Calogero case.

4. CONCLUSION

In conclusion,
a generalisation of the Calogero model has been proposed
which presents some analogy
with the way the anyon model is extended to a general
topological quantum model [8]. The similarities  are patent when the
asymptotic behavior at coinciding points, and the analytical form of the ground
state
and
of some excited states, are considered.
Unfortunately,
the lack of an explicit mapping between the $N$-body Calogero
model and the
$N$-anyon model in the lowest Landau level (only in the $2$-body case
one has been able to construct an explicit mapping [10]) did not allow
for a direct proof of this correspondance, and thus for a systematic
way to build the entire spectrum.
On the other hand, the algebraic approch displayed in [1] has been useful
to construct some
excited states.
It would certainly be quite interesting to find a way towards a complete
resolution of this multi-species Calogero model.

\vfill\eject

REFERENCES :

1) L. Brink, T.H. Hansson, S. Konstein and M.A. Vasiliev, Nucl. Phys.
B401 (1993) 591  ; T.H. Hansson, J. M. Leinaas and J. Myrheim, Nucl.
   Phys. B384 (1992) 559 ; A. P. Polychronakos, Phys. Rev. Lett. 69
(1992) 7703

2) F. Calogero, J. Math. Phys. 10 (1969) 2191, 2197 ; 12 (1971) 419

3) J. M. Leinaas and J. Myrheim, Nuovo Cimento   B37 (1977) 1

4) A. Dasnieres de Veigy and S. Ouvry, Phys. Rev. Lett. 72 (1994) 600

5) S. Isakov, Int. Jour. Mod. Phys. A9 (1994) 2563;
Int. Jour. Mod. Phys. B8 (1994) 319

6) Y-S Wu, Utah Preprint (1994); D. Bernard and Y-S Wu, Saclay Preprint
(1994); M.V.N. Murthy and R. Shankar, IMSc Preprint (1994)

7) F. D. M. Haldane, Phys. Rev. Lett. 67 (1991) 937

8) A. Dasnieres de Veigy and S. Ouvry, Phys. Lett. B307 (1993) 91

9) A. Comtet, J. McCabe and S. Ouvry, Phys. Lett. B260 (1991) 372

10) J.M. Leinaas, private communication.

\end{document}